\documentclass[a4paper,usenatbib]{mnras}

\usepackage{savesym}
\usepackage{amsmath}
\savesymbol{iint}
\savesymbol{iiint}
\usepackage{txfonts}
\restoresymbol{TXF}{iint}
\restoresymbol{TXF}{iiint}

\usepackage{graphicx}	
\usepackage[pdftex]{epsfig}

\newcommand{\xte}{{\it RXTE}}
\newcommand{\eps}{{\rm ergs\,s^{-1}}}

\newcommand{\src}{Aql~X-1}
\newcommand{\aqlexp}{1850}
\newcommand{\aqldiprate}{$0.10_{-0.05}^{+0.07}$~cycle$^{-1}$}
\newcommand{\altdipratelim}{$4.4\times10^{-3}$~cycle$^{-1}$}
\newcommand{\bestdipratelim}{$1.4\times10^{-3}$~cycle$^{-1}$}
\newcommand{\aqlincl}{72--$79^\circ$}
\newcommand{\ucbexp}{1500}  
\newcommand{\ucborbit}{2186}
\newcommand{\nsource}{24}

\title[Intermittent dipping in a LMXB]{Intermittent dipping in a low-mass X-ray Binary}

\author[D. K. Galloway et al.]{Duncan K. Galloway,$^{1,2}$\thanks{E-mail: duncan.galloway@monash.edu}
Alishan N. Ajamyan,$^{1}$
James Upjohn$^{1}$ \newauthor
and Matthew Stuart$^{3}$
\\
$^{1}$School of Physics \& Astronomy, Monash
  University, VIC 3800, Australia\\
$^{2}$also Monash Centre for Astrophysics, Monash
  University, VIC 3800, Australia\\
$^{3}$School of Mathematical Sciences, Monash
  University, VIC 3800, Australia
}

\date{Accepted 2016 June 29. Received 2016 June 29; in original form 2016 May 30}

\pubyear{2016}

\begin{document}
\label{firstpage}
\pagerange{\pageref{firstpage}--\pageref{lastpage}}
\maketitle

\begin{abstract}
Periodic dips observed in $\approx20$\% of low-mass X-ray binaries are thought to arise from obscuration of the neutron star by 
the outer edge of the accretion disk.
We report the detection with the {\it Rossi X-ray Timing Explorer}\/ of two dipping episodes in \src, not previously a known dipper. The X-ray spectrum during the dips 
exhibited an elevated neutral column density, by a factor between 1 and almost two orders of magnitude. Dips were not observed in every cycle of the 18.95-hr orbit, so that the estimated frequency for these  events is  \aqldiprate.
This is the first confirmed example of intermittent dipping in such a system.
Assuming that the dips in \src\/ occur because the system inclination is intermediate between the non-dipping and dipping sources, implies a range of \aqlincl\ for the source. This result lends support for the presence of a massive ($>2\ M_\odot$) neutron star in \src, and further implies that
$\approx30$ additional LMXBs may  have inclinations within this range, 
raising the possibility of intermittent dips in those systems also. 
Thus, we searched for dips from \nsource\ other bursting systems, without success. For the system with the largest 
number of dip phases covered, 4U~1820$-$303, 
the nondetection implies a 95\% upper limit to the dip frequency of \bestdipratelim. 
\end{abstract}


\section{Introduction}

Periodic, irregular dips in the X-ray intensity of low-mass X-ray binaries (LMXBs) were first observed in the early 1980s \cite[]{wm85}.
The dips are 
generally attributed to partial obscuration of the neutron star by a thickened region of the accretion disk close to the line joining the two centres of mass \cite[e.g.][]{dt06}.
The dips are typically irregular in shape and depth, with duty cycle between 10--30\%, and (in most systems) are accompanied by an increase in the absorption column density.

XB~1916$-$050 was the first such example discovered \cite[]{white82,walter82},
with dips recurring at the $\approx50$~min orbital period.
\cite{wm85} described 4 systems showing periodic dips,
and one additional system (X~1624$-$490) in which the dips were not yet known to be periodic.
Since then, the list of known dippers has grown to include
EXO~0748$-$676, 4U~1254$-$69 (XB~1254$-$690), MXB~1659$-$298,  4U~1746$-$371, 4U~1323$-$62, XTE~J1710$-$281, GRS~1747$-$312, and possibly also XTE~J1759$-$220 and 1A~1744$-$361,  \cite[e.g.][]{lmxb07}.
The most recent discovery of dipping behaviour is in the 24.27-d binary and burst source, GX~13+1 \cite[]{iaria14}.

The dips are characterised by a decrease in X-ray intensity, and (usually) an increase in spectral hardness, arising from additional absorption at the low-energy ($< 10$~keV) part of the X-ray spectrum.
The lack of photoabsorption in shallow dips seen from X~1755$-$338 was explained by metal-poor obscuring material; the required abundances are  1/600 of the solar value. Similarly, a factor 10--60 shortfall in degree of photoelectric absorption was also noted for XB~1916$-$50 \cite[]{wm85}.

Dips may occur over $\approx30$\% of the orbital cycle, generally just prior to inferior conjunction, and  culminating in some sources (EXO~0748$-$676 and MXB~1659$-$298) in an eclipse \cite[e.g.][]{parmar86,cw84}.
The inclination angle for dipping sources is thought to be higher than for the non-dippers, and for the systems which also show eclipses is higher again \cite[e.g.][]{motch87}.

Conventionally, the sample of LMXBs was clearly divided into dippers (which showed a dip in almost every orbital cycle) and non-dippers, which never exhibited dips. 
A possible exception was the candidate absorption event  reported from the ultracompact binary 4U~1820$-$303 \cite[]{csb85} following a search for high hardness ratios with HEAO A-2. Scanning observations of the source were made between 1977 August and 1978 March; one of the scans, on 1977 Sep 27, 23:33:56 UT, was found with hardness ratio increased compared to previous scans by a factor of almost 3.
The neutral column density was inferred to have increased for the ``abnormal'' scan. Because of the low duty cycle of the scanning observations, the duration of the event was not well constrained \cite[$\leq2$~hr;][]{mrg88},) although the variation was seemingly not related to the (energy-independent) orbital X-ray intensity modulation.

Recently, a single dipping event was observed by the {\it Rossi X-ray Timing Explorer}\/ ({\it RXTE})  from the 18.95-hr binary \src\  \cite[]{gal12b}, not previously known as a dipping source.
\src\ is one of the most prolific X-ray transients known, exhibiting bright ($L_X\approx10^{37}\ \eps$) outbursts every few hundred days since 1969 \cite[e.g.][]{campana13}. The transient outburst profiles are quite variable, with a range of durations and fluences noted by several authors \cite[e.g.][]{gungor14}. In recent years, the source has exhibited ``long high'' outbursts in 2011 and 2013, and weaker outbursts in each of the following years \cite[e.g.][]{waterhouse16}.

Here we describe in more details the properties of the event observed from \src, and also report on a search for additional dipping events in a large sample of LMXBs.

\section{Observations and Analysis}
\label{analysis}

We analysed {\it Rossi X-ray Timing Explorer}\/ ({\it RXTE};) observations of low-mass X-ray binaries assembled for the Multi-INstrument Burst ARchive (MINBAR\footnote{\url{http://burst.sci.monash.edu/minbar}}).
{\it RXTE}, operational from its launch on 1995 December 30 through to the end-of-mission on 2012 January, consists of three instruments sensitive to cosmic X-rays: the All-Sky Monitor (ASM), the High-Energy X-ray Timing Experiment (HEXTE), and the Proportional Counter Array (PCA). The PCA is comprised of five identical Proportional Counter Units (PCUs), sensitive to photons in the 2--60~keV energy range, and together presenting an effective area of approximately 6500~cm$^2$ \cite[]{xte96}. The field-of-view is approximately circular with a radius of 1~degree, determined by a passive hexagonal-lattice collimator, and the instrument offers no spatial resolution.
Photons detected by the PCA are time-tagged to a precision of approximately $1\mu$s, and processed by an array of on-board event analysers (EAs). Two of the EAs are dedicated to producing standard-mode analysis results; Standard-1 data offers no spectral resolution but 0.125~s binned lightcurves, while Standard-2 data offers spectra accumulated every 16~s within 129 spectral channels covering the energy range.

The MINBAR project builds on an earlier sample of bursts observed by \xte\/ \cite[]{bcatalog}, and seeks to accumulate all observations of burst sources by \xte\/ as well as {\it BeppoSAX}/WFC and {\it INTEGRAL}/JEM-X. 
As part of the data analysis pipeline for the MINBAR observations, we calculated soft and hard X-ray colours from the background-subtracted Standard-2 lightcurves, as the ratio of counts in the 3.7--4.9~keV and 2--3.7~keV bands, and the 8.6--17.7~keV and 4.9--8.6~keV bands, respectively. In contrast to the analysis of \cite{bcatalog}, we made no correction for the gain changes throughout the lifetime of the instrument, as we focused on searches for dipping events usually within a single $\approx90$~min orbit.

The dipping activity in Aql~X-1 during the observation on 2011 October 21 (MJD 53715) was discovered serendipitously via inspection of the lightcurve around the time of a thermonuclear burst \cite[][ see also \S\ref{results}]{gal12b}.
The dips
were characterised by
both  soft and hard colours significantly in excess of the
typical measurements before and after, in addition to a decrease in intensity.
We subsequently used this event as a template to perform an exhaustive search in this  and  other systems with a substantial accumulated exposure with \xte, as follows.

First, we filtered the colour measurements to exclude intervals with high variance, including unphysically large positive or negative values, usually attributable to low source intensity, e.g. at the end of a transient outburst.

Second, we divided the available data from each source into contiguous segments, in which the maximum allowed
gap between segments was 0.006~d. 
We further selected only data segments with at least 10 data points,
as it would not be possible to confirm a dip in a sparser segment.

From each segment we calculated
the mean $\bar{C}_b$ and standard deviation $\sigma_b$ for each band $b$ (hard and soft colours), and identified candidate events as high-significance outliers from the average distribution. We estimated the significance as
\begin{equation}
    S_{b,i} = \frac{C_{b,i}-\bar{C}_b}{\sigma_b}
\end{equation}
and inspected only those measurements where $S_{b,i}$ was in excess of a threshold established empirically from the analysis of the Aql~X-1 data, of $S_{b,i}>5$.

Finally, we inspected the intensity data covering the 
dip candidate, to test for the coincident decrease in intensity expected for a genuine dipping event.

\section{Results}
\label{results}

\begin{figure}
 \includegraphics[width=\columnwidth]{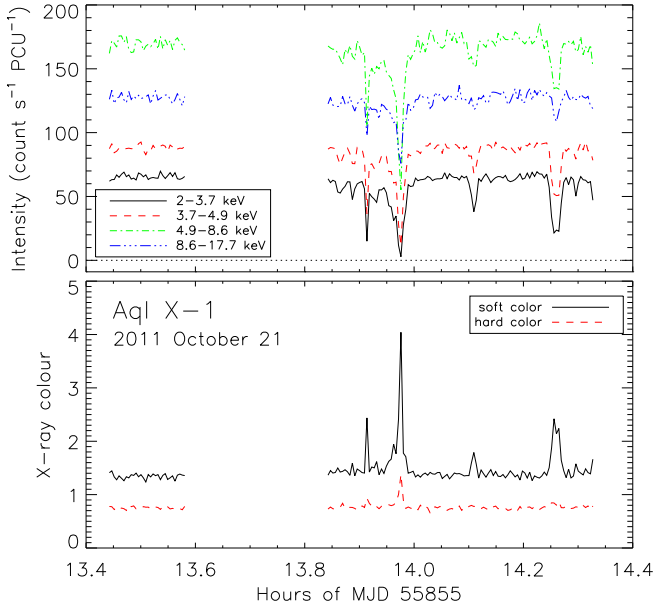}
\caption[]{X-ray intensity of Aql X-1 on 2011 October 21 as a function of energy, calculated from 16-s binned Standard-2 mode data. The X-ray intensity in the 2-3.7 keV, 3.7-4.9 keV, 4.9-8.6 keV and 8.6-17.7 keV bands (top panel) shows that the dip is deepest at lower energies, producing an overall harder X-ray spectrum. This hardening of the spectrum is confirmed by the soft and hard X-ray colours (calculated from the ratio of 3.7-4.9 keV to 2-3.7 keV intensity, and 8.6-17.7 keV to 4.9-8.6 keV, respectively; bottom panel).
\label{fig:dip2} }
\end{figure}

The first instance of dipping behaviour detected from 
\src\ was in an observation by \xte\/ on 2011 October 21 (obsid 96440-01-02-02), eight days after the detection of a new outburst \cite[]{yamaoka11}. 
Beginning on October 21.579 UT (MJD 55855.583), a series of four dips were observed in the X-ray intensity, lasting between 5--60~s each (Fig. \ref{fig:dip2}). The first two dips were separated by 3.9~min; then 7.8~min to the next two (which occurred as a closely separated pair), and 9~min to the last. The observation ended 3~min later. During the second (and deepest) dip, the count rate was reduced to less than 20\% of the level before or after; the other dips reached between 38--67\% of the count rate before or after. One proportional counter unit (PCU \#2) was operating, and the background-subtracted 2--60~keV count rate outside the dips was typically 470~counts~s$^{-1}$.

The fractional decrease in intensity during the dips was greatest at low energies (Fig. \ref{fig:dip2}), resulting in a hardening of the X-ray spectrum. Both the soft and hard X-ray colours were significantly higher compared to the non-dip data. During the second (and deepest) dip, the soft (hard) colour reached 4.0 (1.39), compared to 1.3 (0.8) outside the dips. 
These values were discrepant from the mean colours for the observation at an estimated significance of 8.3 (6.4)~$\sigma$ (Table \ref{tab:aqlcand}).

\subsection{Search for additional dips from \src}
\label{moredips}

We carried out a search (as described in \S\ref{analysis}) for similar events in the entire sample of \xte\/ observations of \src, totalling 1.85~Ms (Table \ref{tab:src}). One additional candidate was found, on 2005 December 11 (MJD 53715.173), almost 6 years before the 2011 October event. 
This observation fell approximately one month after the start of the most recent outburst, which likely began on 2005 November 12 \cite[]{rodriguez05}.
Only one interval with substantial ($\approx40$\%) reduction in intensity was observed in 2005 December, lasting approximately 34~s. The dip onset was gradual, but the recovery much steeper, similar in profile to the deepest dipping event observed in 2011 October.
Four PCUs were active during the observation, but the intensity was lower than for the 2011 October event, by approximately a factor of two.
The significance of the maximum deviation from the colour measurements during the observation was similar in both cases, at 6--8$\sigma$
 (Table \ref{tab:aqlcand}).

We also list in Table \ref{tab:aqlcand} the orbital phase
range of the dip activity, based on the ephemeris for the 18.95-hr orbit adopted by \cite{wry00}. 
We calculated the phase $\varphi_i$ of each data point at time $t_i$, via
\begin{equation}
     \varphi_i = \left(\frac{t_i - T_0}{P_{\rm orb}}\right) \mod 1 
\end{equation} 
where $t_i$ is the time of each data point, $ P_{\rm orb}$ the orbital period, and $T_0$ the reference time (inferior conjunction). Both $ P_{\rm orb} $ and $ T_0 $ have associated uncertainties, so the calculated orbital phase value will also have an  uncertainty which grows with time elapsed since the ephemeris epoch:
\begin{equation}
u(\varphi) = \frac{1}{P_{\rm orb}}\sqrt{u(T_0)^{2} +  N_i^2u(P_{\rm orb})^{2}}
\end{equation}
where $u(A)$ is the uncertainty in quantity $A$, and $N_i=(T_0-t_i)/P_{\rm orb}$ is the number of orbital cycles that has elapsed between the reference epoch $T_0$ and the observation time $t_i$.
For both events observed in \src,
the phase is consistent with the range of orbital phases where dipping is most common for LMXBs (taking into account the uncertainty). 
We estimate the probability of such an agreement by chance as approximately 2.6\%.

We calculated the equivalent exposure time for dips in the \xte\/ sample as the (fractional) number of orbital cycles for which the entire phase range of interest (0.7--1.0) would be covered. For \src, we calculate the total exposure in this phase range as 26.7~cycles. We then estimated the average likelihood of dips for \src\ as 2 per 26.7 equivalent cycles covered, or an overall dip frequency of roughly $ 2/26.7 = 0.075$~cycle$^{-1}$. 
By assuming that the number of dips detected is Poisson-distributed, we can more precisely estimate the likely range of the dip occurrence frequency of
\aqldiprate.
In practise, the data covering each dip phase may be incomplete, interrupted by (satellite) orbital data gaps and scheduled observations of other sources. We may have missed additional dips in those data gaps, due to incomplete coverage of the dip phase. Thus, the dip occurrence frequency calculated above must be considered a lower limit to the actual dip frequency.

\begin{table}
\caption{Aql X-1 dip candidates and their associated properties
 \label{tab:aqlcand}}
\begin{tabular}{clccc}
$T_{\rm cand}$ (MJD)
  & obsid
  & $\sigma_{\rm soft}$ 
  & $\sigma_{\rm hard}$ 
  & $ \varphi $ \\
\hline
53715.173(1) & 91414-01-08-00 & 8.3 & 6.4 & $0.92\pm0.06$ \\ 
55855.583(1) & 96440-01-02-02 & 7.7 & 8.2 & $0.03 \pm0.09$ \\ 
\hline
\end{tabular}
\end{table}

\subsection{Spectral analysis of dip candidates in Aql X-1}

Here we compare the X-ray spectrum measured by the PCA in the dip intervals, to the rest of each observation.
For the observation on 2011 Oct 21 (ID 96440-01-02-02), only PCU \#2 was operational, and we found the best fit to the observation-averaged spectrum with a model consisting of a Comptonisation continuum \cite[{\tt compTT} in {\sc XSpec};][]{tit94} also with a Gaussian component to model the Fe K$\alpha$ fluorescence line around 6.4~keV. The model included the effects of absorption by neutral material along the line-of-sight, with column density frozen at $n_H=4\times10^{21}\ \mathrm{cm}^{-2}$ \cite[]{campana03b}. The assumed systematic error, following the recommendations of the instrument team, was 0.5\%.

\begin{figure}
 \includegraphics[width=\columnwidth]{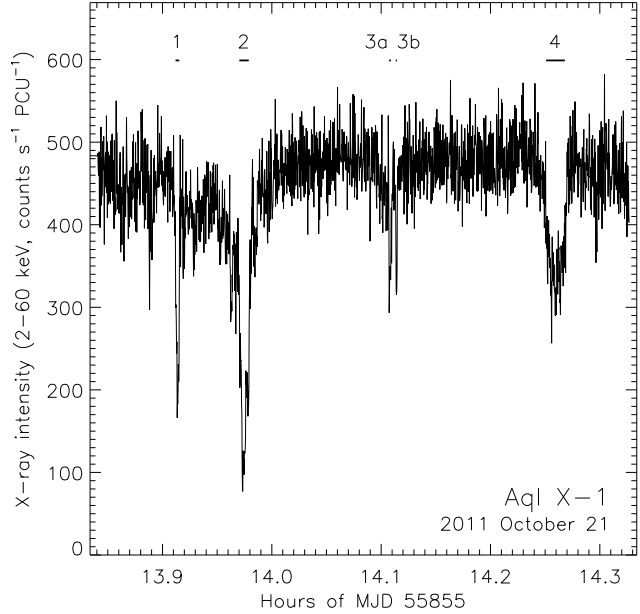}
 \caption[]{X-ray intensity of Aql X-1 during the RXTE observation on 2011 October 21 (obsid 96440-01-02-02). Shown is the 1-s binned lightcurve of the Standard-1 mode data, which includes photons in the energy range 2-60 keV. Note the multiple deep dip features, numbered 1, 2, 3a, 3b and 4; the thick black lines indicate the extent of spectral extraction for each dip. The lightcurve has been background subtracted and corrected to the solar system barycentre.
\label{fig:dip1} }
\end{figure}

We subsequently divided up the observation into five separate dip segments (labeled 1, 2, 3a, 3b, and 4;  Fig. \ref{fig:dip1}), and extracted a spectrum for each of these segments, as well as a spectrum {\it excluding}\/ all of the dip intervals, for comparison. Because the time resolution for Standard-2 data is 16~s, and the dip intervals identified did not correspond exactly to the start and end times of the 16-s bins, we extracted 64-channel spectra from Event-mode data, and calculated spectral responses and background appropriately. We then performed a joint fit of the non-dip and dip spectra, with a spectral model based on the best-fit model for the average, and experimented with freeing joint parameters between the spectra to determine the best way to model the variations during the dips.

We found that freeing the Comptonisation component normalisation alone, while leaving the $n_H$ fixed at $4\times10^{21}\ \mathrm{cm}^{-2}$,  did not give a satisfactory fit ($\chi^2_\nu=13.97$ for 169 degrees-of-freedom). A similar result was obtained if the Gaussian normalisation was also thawed between the dips ($\chi^2_\nu=13.71$).
Conversely, a fit where the neutral column density $n_H$ alone was free to vary between the dip and non-dip spectra, also resulted in a statistically unacceptable fit ($\chi^2_\nu=2.89$).

Allowing both the Comptonisation component normalisation, and the $n_H$ column density to vary between the dips (while leaving $n_H$ frozen at $4\times10^{21}\ \mathrm{cm}^{-2}$ for the non-dip spectrum) gave a much better fit, with $\chi^2_\nu=230.55/165=1.397$. 
The largest fitted column density was $(27.9\pm1.2)\times10^{22}\ \mathrm{cm^{-2}}$ for interval 2, with the other dips exhibiting fitted column densities between (6--$19)\times10^{22}\ \mathrm{cm^{-2}}$.

We found similar results for the single dip observed in the 2005 December observation. Here we extracted a single spectrum for the sole dipping episode (labeled as ``Interval 1'' in Table \ref{tab:aqlspec}), with duration 32~s, and a comparison spectrum excluding this interval.
As before, a {\tt compTT} continuum component with a Gaussian simulating an Fe K$\alpha$ emission line provided the best overall fit, with the $n_H$ value during the dip significantly in excess of the non-dip value, at $(14.0\pm0.5)\times10^{22}\ \mathrm{cm^{-2}}$. However, the scattering electron temperature $kT_e$ could not be constrained. This did not affect the overall quality of the fit, 
but did result in a disproportionately lower value of the {\tt compTT} normalisation, because the integrated flux of this component also depends on the other spectral parameters via the scattering temperature and optical depth (note the absence of a prefactor for the $K_{\rm compTT}$ parameter for the 2005 December fits in Table \ref{tab:aqlspec}).  The combined fit of the dip and non-dip spectra achieved a $\chi^2$ fit statistic of 133.2 for 101 degrees of freedom, indicating a statistically acceptable fit.

\begin{figure}
\includegraphics[height=\columnwidth,angle=270]{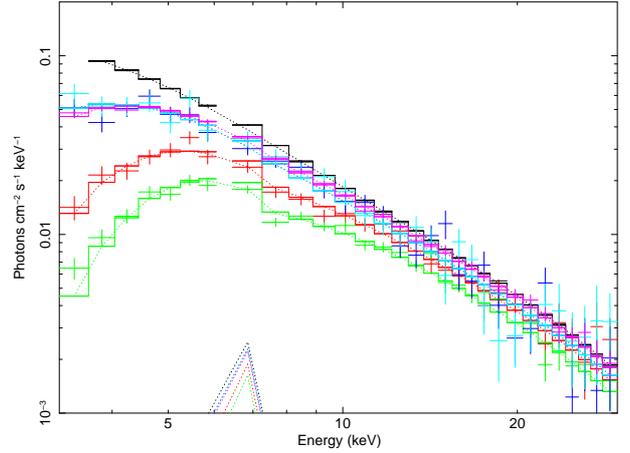}
\caption[]{X-ray spectra of Aql X-1 during the dipping event of 2011 October 21. Shown are the 
PCU \#2 spectra ({\it crosses}) and best-fit model curves ({\it solid histograms}) for each of the non-dip {\it black}, and dip spectra. Each of the dip intervals 1, 2, 3a, 3b and 4 are plotted with a different colour: red, green, blue, aqua, and magenta, respectively. Note the disproportionate reduction in the intensity below $\approx6$~keV for the dip spectra, most notable for the dip 2 interval ({\it green}).
\label{fig:dip4} }
\end{figure}

\begin{table*}
\caption{Spectral characteristics of Aql X-1 data during the 
dipping episodes
in 2005 December 11 and 2011 October 21.
 \label{tab:aqlspec}}
\begin{tabular}{lcccccc}
 & & \multicolumn{5}{c}{Interval} \\
Parameter & units & non-dip & 1 & 2 & 3a \& 3b & 4 \\
\hline 
\multicolumn{7}{c}{2005 December 11} \\
\hline
Time range & s from start & 0--3420 excluding dips & 1783--1815  \\
$n_H$ & $10^{22}\ \mathrm{cm}^2$  & (0.4) & $14.0\pm0.5$  \\
$E_\mathrm{Fe}$ & keV  & (6.4) \\
$\sigma_\mathrm{Fe}$ & keV & $1.09_{-0.07}^{+0.06}$ \\
$K_\mathrm{Fe}$ & $10^{-3}$ photons cm$^{-2}$ s$^{-1}$ & $5.0\pm0.4$\\
$T_0$ & keV & $0.755\pm0.009$ \\
$kT$ & keV & $>124$  \\
$\tau$ & & $(1.58_{-0.04}^{+1.80})\times10^{-2}$ \\
$K_\mathrm{compTT}$ & $10^{-3}$ & $0.886_{-0.007}^{+0.401}$ & $0.744_{-0.011}^{+0.338}$  \\
$\chi^2$ (dof) & & 133.2(101) \\
\hline
\multicolumn{7}{c}{2011 October 21} \\
\hline
Time range & s from start & 0--3420 excluding dips & 1779--1791 & 1989--2020 & 2482--2488, 2504--2508 & 2999--3061 \\
$n_H$ & $10^{22}\ \mathrm{cm}^2$ &  (0.4) & $18.7\pm1.2$ & $27.9\pm1.2$ & $6.0\pm0.8$ & $8.0\pm0.3$ \\
$E_\mathrm{Fe}$ & keV  & $6.60\pm0.10$ \\
$\sigma_\mathrm{Fe}$ & keV  & (0.5) \\
$K_\mathrm{Fe}$ & $10^{-3}$ photons cm$^{-2}$ s$^{-1}$ & $3.9\pm0.6$\\
$T_0$ & keV  & $0.894\pm0.013$ \\
$kT$ & keV  & $8.83^{+0.42}_{-0.36}$ \\
$\tau$ & &  $3.13\pm0.11$ \\
$K_\mathrm{compTT}$ & &  $0.101\pm0.005$ & $0.084\pm0.005$ & $0.073\pm0.004$ & $0.089\pm0.005$ & $0.098\pm0.005$ \\
$\chi^2$ (dof) &  & 230.6 (179) \\
\hline
\end{tabular}
\end{table*}

\subsection{Search for dips from other LMXBs}

We repeated the dip search for non-dipping sources\footnote{See \url{http://burst.sci.monash.edu/sources}} for which we had data in-hand as part of the MINBAR sample (Table \ref{tab:src}). 
We prioritised the sources based on the total exposure accumulated with \xte\/ over the mission, and terminated our search when the estimated number of dip phases observed dropped below 1 (see \S\ref{kfac}).
No plausible dip candidates were detected.
Ideally such a search would also be extended to the non-bursting LMXBs, but analysis data for those sources were not available. Since the burster sample represents approximately half the known LMXB population \cite[e.g.][]{lmxb07}, we might expect at most one additional intermittent dipper to be detectable. 

The best candidate for dips in sources other than \src\ was in
Cyg X-2, which  exhibited in a few observations a  pattern of intensity decreases, similar to that observed during dips. 
During some (but not all) of these intervals of lower intensity, the soft and hard colours 
were elevated, but not significantly.
Spectral analysis of the best candidate dip interval, in observation 70016-01-01-00, exhibited no significant increase in the column density. Thus, we expect that the intensity variations may be due to other causes than dips.

As for \src, for those systems with measured orbital ephemerides, we summed the total exposure in the \xte\/ sample over the phase range in which dips are expected.
4U~1636$-$536 had the largest total exposure, with 682 total orbit equivalents. 
With no dips detected, we can calculate an upper limit on the dip rate, of 
\altdipratelim (95\%). 
This is more than an order of magnitude lower than the estimated frequency for \src\ (\S\ref{moredips}), indicating that we can rule out intermittent dipping behaviour conclusively in this system.
Despite the smaller total exposure compared to 4U~1636$-$536, an even larger number of dip phases was observed for the ultracompact binary 4U~1820$-$303, 
due to the  11.46~min orbit.
The total number of complete dip phases covered was \ucborbit, leading to an upper limit on the dip rate of \bestdipratelim\ (95\%).
This limit is difficult to reconcile with the claim of a dipping event by \cite{csb85}.

\begin{table*}
\caption{Low-mass X-ray binaries for which dip searches were carried out
 \label{tab:src}}
\begin{tabular}{lccccl}
  Source name
  & R.A.
  & Dec.
  & $P_{\rm orb}$ (hr)
  & Tot. exp (ks)
    & Ref. \\
\hline 
Aql X-1 & 19 11 16.05 & +00 35 05.8 & 18.95 & \aqlexp & \cite{wry00} \\
\hline
4U 1636$-$536 & 16 40 55.50 & $-53$ 45 5.0 & 3.8 & 4270 & \cite{casares06} \\
XTE J1701$-$462 & 14 00 58.50 & $-46$ 11 8.60 &  & 2730 &  \\
4U 0614+09 & 6 17 7.3 & +09 8 13 & 0.50--0.83? & 2680 & \cite{baglio14,madej13} \\
Cyg X-2 & 21 44 41.2 & +38 19 18.0 & 236.2 & 2230 & \cite{premach16} \\
4U 1608$-$522 & 16 12 43.05 & $-52$ 25 23 & 12.89 & 2080 & \cite{wachter02} \\
4U 1728$-$34 & 17 31 57.5 & $-33$ 50 5 & 0.18? & 1680 & \cite{gal10b} \\
4U 1820$-$303 & 18 23 40.5 & $-30$ 21 40.1 & 0.19 & \ucbexp & \cite{anderson97} \\
4U 0513$-$40 & 05 14 6.6 & $-40$ 2 37 & 0.283 & 1490 & \cite{fiocchi11} \\
SAX J1808.4$-$3658 & 18 08 27.5 & $-36$ 58 44.3 & 2.01365 & 1420 & \cite{chak98d} \\
4U 1705$-$44 & 17 8 54.5 & $-44$ 6 7.4 &  & 1240 &  \\
4U 1702$-$429 & 17 6 15.3 & $-43$ 02 8.7 &  & 1210 &  \\
XTE J1810$-$189 & 18 10 20.7 & $-19$ 04 11 &  & 1150 &  \\
GX 17+2 & 18 16 1.4 & $-14$ 02 11 & & 1120 &  \\
4U 1735$-$444 & 17 38 58.3 & $-44$ 27 00 & 4.65 & 1003 & \cite{casares06} \\
GS 1826$-$24 & 18 29 28.2 & $-23$ 47 29 & 2.088 & 958 & \cite{homer98} \\
HETE J1900.1$-$2455 & 19 00 9.8 & $-24$ 54 4.3 & 1.38757 & 775 & \cite{kaaret05b} \\
2S 0918$-$549 & 09 20 27 & $-55$ 12 24.7 &  & 552 &  \\
4U 1722$-$30 & 17 27 33.2 & $-30$ 48 07 &  & 534 &  \\
GX 3+1 & 17 47 56 & $-26$ 33 49 &  & 518 &  \\
XTE J1710$-$281 & 17 10 12.3 & $-28$ 07 54 & 3.28 & 454 & \cite{jain11} \\
SAX J1748.9$-$2021 & 17 48 53.5 & $-20$ 22 02 &  & 375 &  \\
IGR J17511$-$3057 & 17 51 09 & $-30$ 57 40 & 3.4688 & 366 & \cite{riggio11} \\
4U 1246$-$588 & 12 49 39.6 & $-59$ 05 13.3 &  & 199 &  \\
GRS 1741.9$-$2853 & 17 45 2.3 & $-28$ 54 49.9 &  & 42.3 &  \\
\hline
\end{tabular}
\end{table*}

\section{Discussion}

The intermittent decrease in intensity observed during two outbursts of \src, within the orbital phase range in which dipping behaviour is seen in other LMXBs, and coupled with the evidence for elevated neutral column density, confirms these events as 
arising from the same physical cause as 
the regular dips seen in high-inclination LMXB systems.

These events are the first confirmed ``intermittent'' dips in a LMXB system, although not the first claimed. 
The candidate event reported from HEAO~A-2 data of 4U~1820$-$303 \cite[]{csb85} helped to motivate 
our own search of \ucbexp~ks of \xte\/ observations of this source, comprising \ucborbit\ total orbit equivalents of exposure in the 0.7--1.0 phase range. However, this search resulted in no plausible dip candidates.

Spectral analysis of the dipping events observed in \src\/ indicates that both the neutral column density $n_H$ and the continuum normalisation $K_{\rm compTT}$ vary significantly 
compared to the non-dip spectra. This behaviour is commonly found in (consistently) dipping systems \cite[e.g.][]{smale92,dt06}, and has been attributed to absorption by partially ionised material, possibly with abundances deviating significantly from solar, and/or partial covering of the (possibly extended) emission region.

While the complete sample of \xte\/ observations of Aql X-1 have covered other inferior conjunctions, no additional dipping behaviour has been observed. Based on the total exposure of the orbital phase in which dips are most often observed in other sources, we estimate the average frequency for dips in \src\ as 
\aqldiprate.

We also searched the accumulated \xte\/ data from \nsource\ other systems for intermittent dipping activity, without success. For the system with the largest exposure of the dip phase range, 
4U~1820$-$303, we place an upper limit on the dip frequency of \bestdipratelim\ (95\%).

The \xte\/ sample allows weak constraints to be placed on the duration of the dipping behaviour in \src. 
We inspected observations before and after each dip, selecting those data which fell in the usual dipping phase range of 0.7--1.0. We also searched for data covering the precise phase range of the deepest dipping features, since in other sources these features may persist for multiple orbital periods.
For the 2005 December dip, we had partial coverage of the dipping phase range 3.9~d before and 1.4~d after. This corresponds to 5 cycles before and 2 after, with the data coverage of the range in each case totalling approximately 31\%. We translate these figures to a $\approx30$\% chance of detecting dipping behaviour in each observation, had it been present; since no dips were detected, we can constrain the duration of the dipping behaviour to $<4.3$~d (7 orbital cycles) at $0.3\times0.3\approx0.1$ confidence.
Neither of these observations, however, included the precise orbital phase range in which the 2005 Dec dip was observed (Table \ref{tab:aqlcand}). The closest coverage of that range was found 240~d before and after the dip observation.

Even poorer coverage was achieved for the 2011 October dip. The closest observation covering the 0.7--1.0 phase range before the dip was 388~d earlier; after, 6.9~d (9 cycles). The coverage of the dipping phase range for the latter observations was only 9\%.
The closest observations covering the precise phase range of the deepest dip observed in 2011 October were 684~d earlier, or 8.7~d (11 cycles) after. This latter observation likely provides the most stringent constraint on the duration of the dipping activity from the \xte\/ data.

\subsection{The effect of system inclination}
\label{kfac}

We consider three possible explanations for the apparent dipping behaviour in \src.
First, that the system inclination 
is intermediate between the population of non-dippers and dippers, such that the normal activity of the disk occasionally allows material at the outer edge to cross the line of sight. According to this explanation, any system with an inclination within some range $\theta_1<i<\theta_2$ would exhibit intermittent dipping behaviour, given enough observational data.
To determine the probability distributions of the angles $\theta_1$, $\theta_2$, as well as the additional limit $\theta_3$ above which a source will exhibit eclipses, we make the following assumptions: 1. the fraction of sources in each group is proportional to $(\cos\theta_i-\cos\theta_{i+1})$; 2. the number of sources detected in each category is Poisson-distributed about the expected value; 3. the expected number of intermittent dipping sources detected is lower by a factor of $(1-0.9^k)$, where $k$ is the typical number of dip phase orbit equivalents observed for each source. This latter factor incorporates both the estimated frequency of dips for intermittent sources, as measured in \src, as well as the typical number of dip phases observed. Here we adopt the median \xte\/ exposure for all the burst sources of 55~ks, and the median orbital period, of 4.65~hr, to calculate $k\approx1$.

\begin{figure}
\includegraphics[width=\columnwidth]{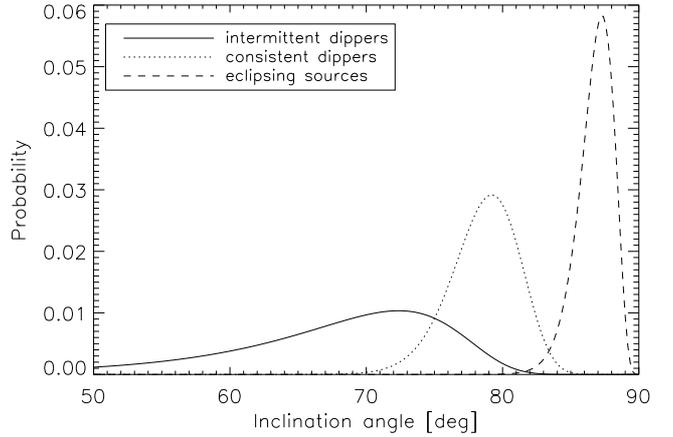}
\caption[]{Inferred probability distributions for angle limits between non-dippers and intermittent dippers $\theta_1$ ({\it solid line}); intermittent and consistent dippers $\theta_2$ ({\it dotted line}); and dippers and eclipsing sources $\theta_3$ ({\it dashed line}). The probability distributions are based on the fraction of dipping and eclipsing sources in the MINBAR sample of bursters.
The overall normalisation scale is arbitrary, but consistent for each angle.
\label{fig:angles} }
\end{figure}

For each angle $\theta_i$ we then marginalise over the other two angles, adopting the observed population of intermittent dipping:consistently dipping:eclipsing sources, of 
1:14:5 (from a total of 106) 
to obtain the probability distributions as plotted in Fig. \ref{fig:angles}.
We find that the most probable range for each angle is 
$\theta_1=72_{-14}^{+ 2}$,
$\theta_2=79.3_{-3.2}^{+1.8}$, and
$\theta_3=87.2_{- 1.8}^{+ 0.7}$
(68\% confidence; central value given has the highest probability of the PDF).
We note that the width of the distribution for $\theta_1$ is quite sensitive to the value adopted for $k$. 
Our analysis suggests that the inclination for \src\ is in the range \aqlincl. Previous estimates of the inclination of \src\ based on optical light curve modelling were confounded by the contribution of light from a nearby unrelated star (\citealt{chevalier99}; see also \citealt{corn07}).
The derived range from our analysis is broadly consistent with the estimate of \cite{garcia99} based on the photometric modulation during outburst, of $\approx70^\circ$. 
Interestingly, the additional evidence for high inclination provided by the detection of dips, lends further support to the case for a massive ($\gtrsim2.7\ M_\odot$) neutron star \cite[]{corn07}.

Since the estimated inclination range for the intermittent dippers is so wide, there are likely $\approx15$ other intermittent dippers in the burster sample, and perhaps the same number in the non-burster LMXB population. That dips have not been detected from any of these $\approx30$ other systems is a consequence of the low dipping frequency, as well as the paucity of data covering the preferred dip orbital phase.

\subsection{Alternative explanations for the intermittent dips}

A second possibility for the presence of intermittent dips in \src\ 
is that the accretion disk in this system is subject to precession, such that the normal vector to the disk plane rotates around the orbital angular momentum axis, bringing the outer edge of the disk into the line-of-sight periodically. Precession in LMXBs is perhaps best known in the pulsar Her~X-1 \cite[e.g.][]{od01}, where a 35-d ``superorbital'' cycle has the disk periodically obscuring the pulsar at it's centre. 
Precession has been suggested to explain quasi-periodic variability of persistent LMXBs \cite[e.g. 4U~1728$-$34,][]{gal03b}, but the relatively subtle effects would likely be indistinguishable from the much higher amplitude variability in transients. 
While our results allow candidate precession periods to be calculated, these are not constrained due to the limited number of events (2). Furthermore, it is not clear if a precession signal would persist over multiple outburst-quiescence cycles.

Third,
the serendipitous observation of dips may be related to changes in the disk structure, perhaps resulting from a change in the disk geometry \cite[signalled by a spectral state transition; e.g.][]{done07}
around the time of the dips. In the  \xte\/ observations subsequent to the 2011 dip, from October 24 onwards, the overall intensity was significantly higher, with soft colour remaining between 1.2-1.3, but the hard colour decreasing to between 0.32-0.4. This increase in intensity, coupled with spectral softening, likely indicates a transition from the ``island'' (hard) spectral state to the ``banana'' (soft) state some time between October 21 and 24. 
However, no such transition was observed around the time of the 2005 dip, nor at anytime during that outburst, which was much weaker than that in 2011.

\section{Conclusions}

We report the detection by \xte\/ of a pair of events in \src\ with properties consistent with the dips found in a subset of low-mass X-ray binaries. These events are the first such ``intermittent'' dips confirmed for a LMXB; given the accumulated exposure on this source, the estimated rate of such events is \aqldiprate. The spectral analysis of the dip segments shows that the inferred neutral column density $n_H$ is elevated typically by an order of magnitude, and up to almost two orders of magnitude, compared to the non-dipping value.
Assuming that the dips arise in \src\ by virtue of a higher-than-average inclination, the implied range is \aqlincl; this is in excess of most other estimates for the source. Given the difficulty of detecting such events in other systems, it is possible that an additional 15 systems may be undetected intermittent dippers. 
Of the possible explanations of the intermittent dipping behaviour, the high inclination seems the most plausible.
We also report on a search of \nsource\ additional systems for dips, which was unsuccessful. For the system with the largest exposure, 4U~1636$-$536, we derive an upper limit on the dip rate of \altdipratelim\ (95\%).

\section*{Acknowledgements}

DKG acknowledges the support of an Australian Research Council Future Fellowship
(project FT0991598).
This paper utilizes preliminary analysis results from the Multi-INstrument Burst ARchive (MINBAR), which is supported under the Australian Academy of Science's Scientific Visits to Europe program, and the Australian Research Council's Discovery Projects 
funding scheme.
This research has made use of data and software provided by the High Energy Astrophysics Science Archive Research Center (HEASARC), which is a service of the Astrophysics Science Division at NASA/GSFC and the High Energy Astrophysics Division of the Smithsonian Astrophysical Observatory.


\clearpage

\bsp	
\label{lastpage}
\end{document}